%% file: main.tex
\documentclass[runningheads]{llncs}
\usepackage{graphicx}
\usepackage{caption}
\usepackage{subcaption}

\usepackage{algorithm}
\usepackage{algpseudocode}
\usepackage{hyperref}
\begin{document}

% \RestyleAlgo{ruled}

% \SetKwComment{Comment}{/* }{ */}
%
\title{Disaggregated Database Management Systems\thanks{This paper appeared in the {\em Performance Evaluation and Benchmarking} - 14th TPC Technology Conference, TPCTC 2022, Sydney, NSW, Australia, September 5, 2022, Revised Selected Papers. Lecture Notes in Computer Science 13860, Springer 2023, ISBN 978-3-031-29575-1.}}
\author{Shahram Ghandeharizadeh$^1$, Philip A. Bernstein$^2$, Dhruba Borthakur$^3$,  Haoyu Huang$^4$, Jai Menon$^5$, Sumit Puri$^6$}
\authorrunning{S. Ghandeharizadeh, et. al.}
\institute{USC, Los Angeles, CA, USA, \email{shahram@usc.edu} \and
Microsoft Research, Redmond, WA, USA, \email{phil.bernstein@microsoft.com} \and
Rockset, San Mateo, CA, USA, \email{dhruba@gmail.com}
\and
Google, Mountain View, CA, USA, \email{haoyuhuang@google.com} \and
Fungible, Santa Clara, CA, USA, \email{jai.menon@fungible.com} \and
Liqid, Broomfield, CO, USA, \email{sumit@liqid.com}
%%%\\ \url{http://www.springer.com/gp/computer-science/lncs} \and ABC Institute, Rupert-Karls-University Heidelberg, Heidelberg, Germany\\ \email{\{abc,lncs\}@uni-heidelberg.de
%}
}
\maketitle              % typeset the header of the contribution
\begin{abstract}
\input{abs.tex}

%%%\keywords{First keyword  \and Second keyword \and Another keyword.}
\end{abstract}
\section{Introduction}\label{sec:intro}
\input{intro.tex}

\section{Hardware Disaggregation}\label{sec:hardware}
\input{hwIntro.tex}

\subsection{Fungible's DPU-Based Disaggregation}
\input{fungible}
\subsection{Liqid's Composable Disaggregated Infrastructure (CDI)}\label{sec:liqid}
\input{liqid}

\section{Memory Disaggregation}\label{sec:memory}
\input{memory}

\section{Disaggregated Database Management Systems}\label{sec:dbms}
\subsection{AlloyDB}\label{ssec:alloydb}
\input{alloydb}

\subsection{Rockset}\label{sec:rockset}
\input{rockset}

\subsection{Nova-LSM}\label{sec:novalsm}
\input{novalsm}

\section{Future Research Directions}\label{sec:future}
\input{future}

\section{Acknowledgments}
We thank Liqid's Bob Brumfield and George Wagner for input on Section~\ref{sec:liqid}.

%
% the environments 'definition', 'lemma', 'proposition', 'corollary',
% 'remark', and 'example' are defined in the LLNCS documentclass as well.
%


\bibliographystyle{plain}
\bibliography{main.bib}

%%%\begin{thebibliography}{8}
%%%\bibitem{ref_article1}
%%%Author, F.: Article title. Journal \textbf{2}(5), 99--110 (2016)

%%%\bibitem{ref_lncs1}
%%%Author, F., Author, S.: Title of a proceedings paper. In: Editor,
%%%F., Editor, S. (eds.) CONFERENCE 2016, LNCS, vol. 9999, pp. 1--13.
%%%Springer, Heidelberg (2016). \doi{10.10007/1234567890}

%%%\bibitem{ref_book1}
%%%Author, F., Author, S., Author, T.: Book title. 2nd edn. Publisher,
%%%Location (1999)

%%%\bibitem{ref_proc1}
%%%Author, A.-B.: Contribution title. In: 9th International Proceedings
%%%on Proceedings, pp. 1--2. Publisher, Location (2010)

%%%\bibitem{ref_url1}
%%%LNCS Homepage, \url{http://www.springer.com/lncs}. Last accessed 4
%%%Oct 2017
%%%\end{thebibliography}
\end{document}

%% file: abs.tex
Modern applications demand high performance and cost efficient database management systems (DBMSs).  Their workloads may be diverse, ranging from online transaction processing to analytics and decision support.  The cloud infrastructure enables disaggregation of monolithic DBMSs into components that facilitate software-hardware co-design.  
This is realized using pools of hardware resources, i.e., CPUs, GPUs, memory, FPGA, NVM, etc., connected using high-speed networks.  
This disaggregation trend is being adopted by cloud DBMSs because hardware re-provisioning can be achieved by simply invoking software APIs. Disaggregated DBMSs separate processing from storage, enabling each to scale elastically and independently.  %They may consist of alternative implementations of a component suitable for different workloads.
They may disaggregate compute usage based on functionality, e.g., compute needed for writes from compute needed for queries and compute needed for compaction.  
They may also use disaggregated memory, e.g., for intermediate results in a shuffle or for remote caching.  
The DBMS monitors the characteristics of a workload and dynamically assembles its components that are most efficient and cost effective for the workload.  
This paper is a summary of a panel session that discussed the capability, challenges, and opportunities of these emerging DBMSs and disaggregated hardware systems.

%% file: intro.tex
Emerging data centers disaggregate hardware into pools of resources and connect them using fast networks such as high-speed Ethernet or Remote Direct Memory Access, RDMA~\cite{haoyu19}.  The pool of resources may include CPUs, GPUs, memory, NVMe, disks, SSDs, FPGAs, and specialized hardware such as Amazon Trainium and Google TPU for machine learning among others.  
The software that implements a database management system may also be disaggregated into micro-services.
Both trends facilitate a software-hardware co-design.
Moreover, they enable composition of micro-services into new services.
For example, an in-memory key-value store may be realized using a subset of micro-services that implement a relational database management system (DBMS)~\cite{disaggregate2019}.  
Assembly of the components must consider the communication latency and employ caches to prevent this latency from dominating performance. 

Disaggregated DBMSs hold the potential to transform today's obsolete practices to enhance efficiency by providing sustainable solutions.
Instead of asking a user to size a server, they may ask a user for their daily budget and desired performance objective.
Now, it is the responsibility of an intelligent agent to assemble the hardware and software to meet the price and performance requirements.
With a high (low) system load, the assembled system may scale-up (down).
The system may use alternative forms of storage that provide different price/performance characteristics~\cite{hierarchy2018}.

\begin{figure}
\centering
\begin{subfigure}[t]{0.48\linewidth}
  \centering
  \includegraphics[width=\linewidth]{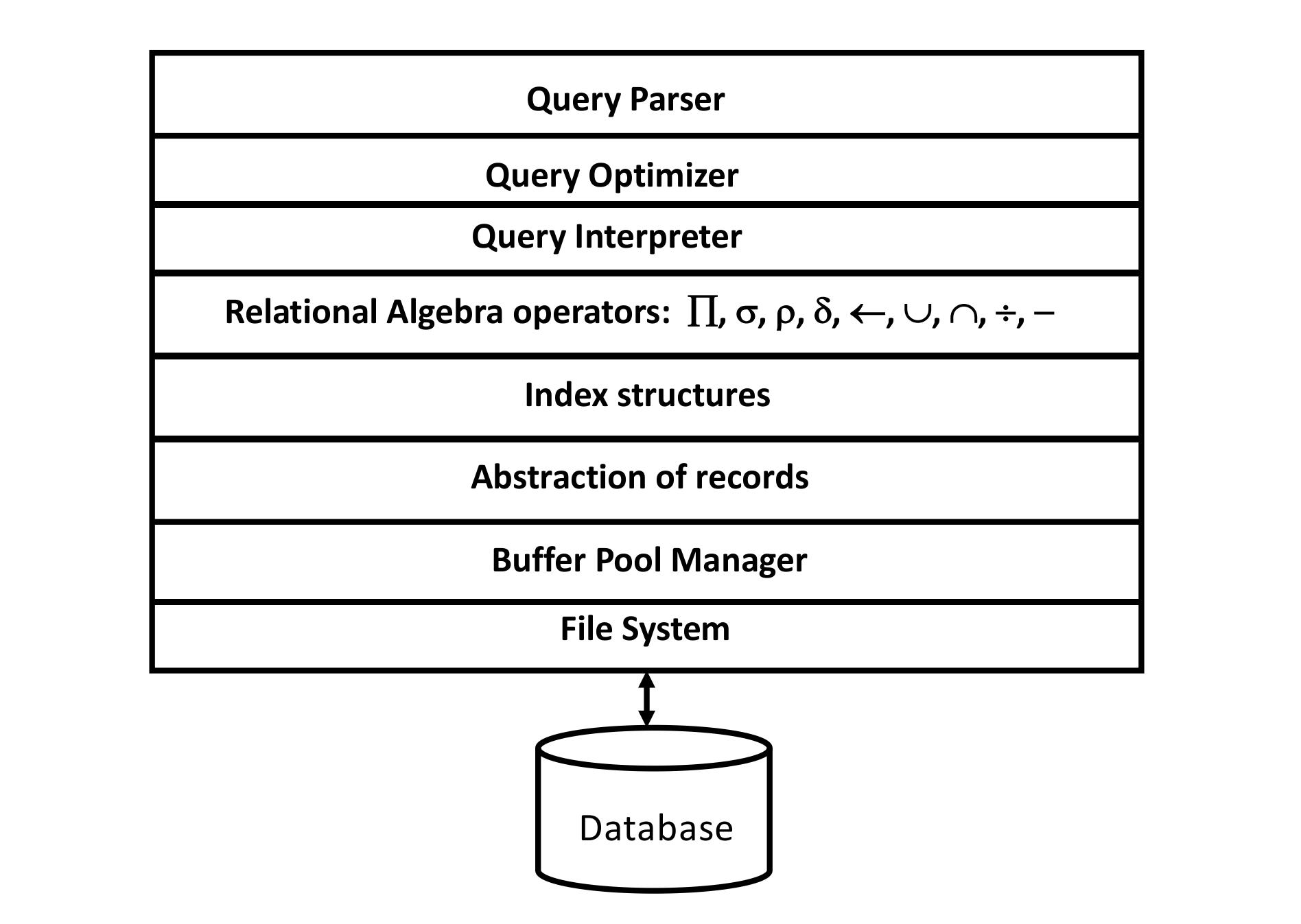}%pdf}
  \caption{A monolithic DBMS.}
  \label{fig:monolithic-db}
\end{subfigure}
% \quad
\begin{subfigure}[t]{0.48\linewidth}
  \centering
  \includegraphics[width=\linewidth]{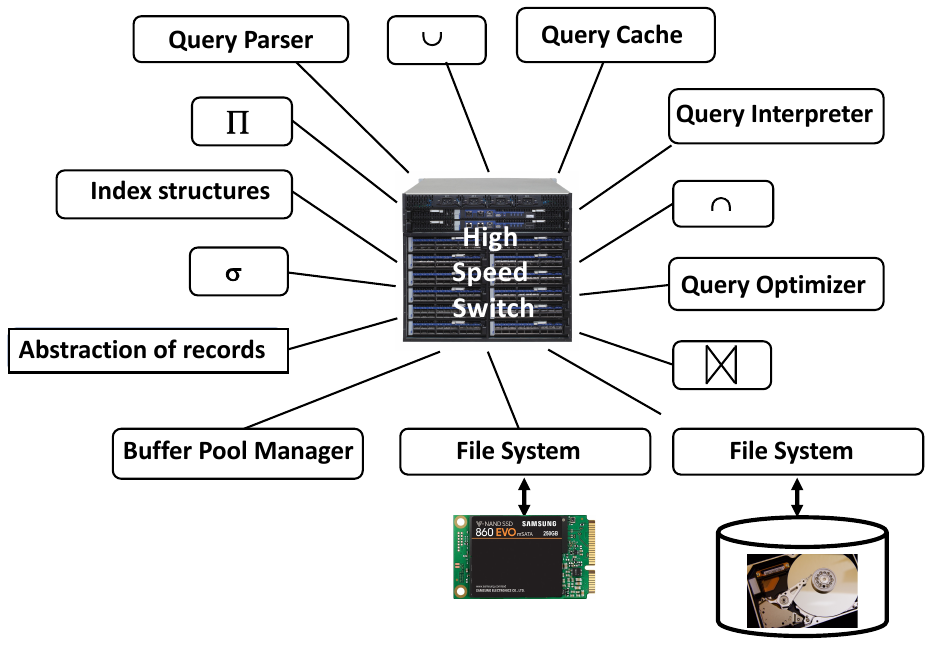}
  \caption{A disaggregated DBMS.}
  \label{fig:nova}
\end{subfigure}
\caption{Today's monolithic DBMS and the envisioned disaggregated DBMS.}
\label{fig:nova-arch}
\end{figure}

Physical data design is a task performed by data administrators to enhance efficiency and meet the performance requirements of an application.  They make decisions such as whether the data is stored in a column format or a row format.  
However, many startups do not have a budget to pay these experts, resulting in inefficiencies.
A disaggregated DBMS will address this by monitoring the system workload and fine-tuning the physical design of the database, choice of hardware including a storage hierarchy, and microservices assembled to realize the most cost effective deployment.

Today's use of data centers by scientists, including those in the area of machine learning and AI, requires them to upload their data to a data center and run their computation in the cloud.  A key question is what data to copy to the cloud? 
%And whether their data is already in the cloud?  
As an example, NIH's-National Library of Medicine has 36.4 Petabytes of Genomic sequencing data on two commercial cloud platforms~\cite{presidentialreport}.
How are scientists to discover and use this data?
Scientists want to know what is the minimum cost configuration for an experiment.  
They may have a fixed budget for running an experiment.  Once the budget is exhausted, they may want to save their result files and have the service shut-down so there are no additional charges~\cite{presidentialreport}.  
They may also want to take a snapshot of their mid-flight experiment that enables them to continue where it was stopped.

A vision for a future system is one that reduces data movement and replication~\cite{presidentialreport}.
One way to realize this is to extend the disaggregation beyond a data center to include a scientist's desktop.  It provides for physical data independence, a concept pioneered by the database community, where the scientist is no longer burdened with the placement of data.  It is the responsibility of the infrastructure to manage placement of data and computation seamlessly.
%This concept of physical data independence may be applied to multiple data centers and many users whose desktops contributes resources.  The processing and data placement may also be performed at the edge using the desktops.% with the final destination identified by the scientist.  %Typically, the default is the scientist's desktop.

The rest of this paper is organized as follows.
Sections~\ref{sec:hardware},~\ref{sec:memory} and~\ref{sec:dbms} describe hardware, memory, and DBMS disaggregation in turn.
Brief future research directions are presented in Section~\ref{sec:future}.

%% file: hwIntro.tex
Storage, GPUs, memory, and other hardware resources are traditionally included as part of a traditional server. 
A limitation of this organization is the box that contains these resources. 
%sheet metal that surrounds resources.
%With GPU, storage and/or FPGA resource requirements, 
Success is effectively limited to what can fit in a box. 
To obtain enough of a critical resource, customers are forced to purchase larger, more expensive servers than required for new deployments or remove and replace a server when that critical resource is maxed-out.

%The downside of this approach is that these resources are inefficiently utilized as they cannot be shared with other servers. Each server must have sufficient hardware resources to deal with the peak system load, and hardware resources are wasted at other times.

Disaggregating the hardware resources from the servers allows for much better utilization of these resources. There are three questions to answer when disaggregating resources:
Which hardware resources are being disaggregated?
What bus, fabric, or network to use to connect the disaggregated resources?
And, what is the performance impact of disaggregation?
%\begin{enumerate}
%    \item Which hardware resources are being disaggregated?
%    \item What bus, fabric, or network to use to connect the disaggregated resources?
%    \item What is the performance impact of disaggregation?
%\end{enumerate}
Consider each question in turn.

\subsubsection{What hardware resources are being disaggregated?}
Storage has been disaggregated for many years. Storage products that disaggregate at the file level (NAS) and at the block level (SAN) have been available for several decades.

GPUs and memory have not been disaggregated until very recently. Products that disaggregate GPUs have become available in the last year. Memory disaggregation is emerging.

\subsubsection{What fabric is used for disaggregation?}
Storage disaggregation has been accomplished using Fiber Channel (FC), Ethernet, and InfiniBand (IB). 
Protocols such as SCSI over FC, NFS over Ethernet and SCSI over Ethernet (iSCSI) have been employed for this purpose. 
An emerging approach is NVMeoF for block level disaggregation. 
See \url{https://nvmexpress.org/wp-content/uploads/NVMe_Over_Fabrics.pdf} for details.
GPU disaggregation has been realized using
%accomplished over 
PCIe Fabrics and over Ethernet.
Memory disaggregation is being attempted over emerging buses such as CXL (\url{https://www.computeexpresslink.org/}).
It is also possible over Ethernet.

\subsubsection{What is the performance impact of disaggregation?}
Disaggregation improves resource utilization, but it may result in 
%comes at the expense of 
some performance loss. 
The closer the performance of the disaggregated resource is to its performance when locally attached to the server, the more likely it is that disaggregation is employed.

There is a tradeoff between using fabrics that allow data center wide disaggregation (such as Ethernet) and those that support disaggregation over shorter distances. It is harder to achieve good performance over data center wide distances.
However, customers like the flexibility offered by disaggregation at scale. 

With new protocols such as NVMeoF, disaggregated storage performance close to that of server attached storage has been demonstrated, even at data center scale. Similarly, new approaches have also shown that disaggregated GPU performance can be anywhere from 80\% to 99\% of local GPU performance at data center scale over Ethernet.

The next two sections present two approaches to hardware disaggregation.

%% file: fungible.tex
A DPU (data processing unit) is a specialized programmable processor tailored to efficiently execute data-centric tasks. These are tasks that require stateful processing simultaneously on multiple high bandwidth streams of data. An increasing fraction of work done in modern data centers is data centric in nature, and CPUs and GPUs are inefficient at such tasks. Disaggregation is essentially a data centric task and DPUs have proven to be very efficient at disaggregation.  As a result, an emerging trend is the use of DPUs for hardware disaggregation.
\begin{figure}[!ht]
    \centering
    \includegraphics[width=0.9\linewidth]{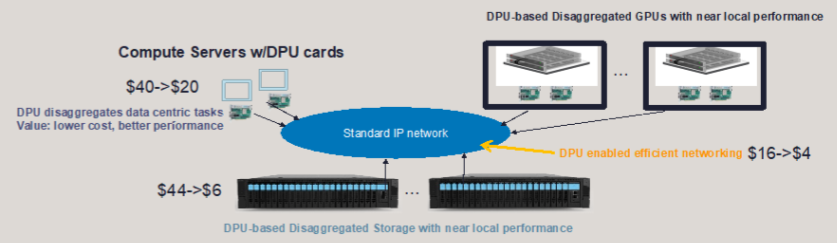}
    \caption{DPU-based disaggregated data center using standard IP networks.}
    \label{fig:dpu}
\end{figure}

Fungible has built 2 DPUs~\cite{dpuwhitepaper},% (\url{https://lp.fungible.com/hubfs/Assets/Whitepapers/The-Fungible-DPU-A-New-Category-of-Microprocessor.pdf}), 
one small enough to fit on a PCIe card inside a server, a second one powerful enough %to use 
to build a disaggregated storage system. 

Using the DPU, Fungible has built a disaggregated storage system with performance indistinguishable from that of server attached storage. This system is called the Fungible Storage Cluster (FSC).  

It has also built a DPU-based PCIe card that plugs into the PCIe slot of a standard server and disaggregates networking, security and storage functions transparently from the server. This allows server cores to be dedicated to running applications instead of being used inefficiently to run infrastructure tasks. 

Finally, it has developed a DPU-based disaggregated GPU appliance with performance between 80-99\% of server attached GPUs. Customers can now run AI/ML on servers without any GPUs, and they have the flexibility to change the mix of CPU cores and GPU cores applied to a given problem.

Fungible’s vision of the next-generation data center is shown in Figure~\ref{fig:dpu}. All servers have DPU based cards. Storage and GPUs are disaggregated and are built with DPUs. The result is a data center that costs about 30\% of one without DPU-based disaggregation. It allows the disaggregated components to be dynamically composed on the fly to meet workload needs for great agility. 

\begin{figure}[!ht]
    \centering
    \includegraphics[width=0.9\linewidth]{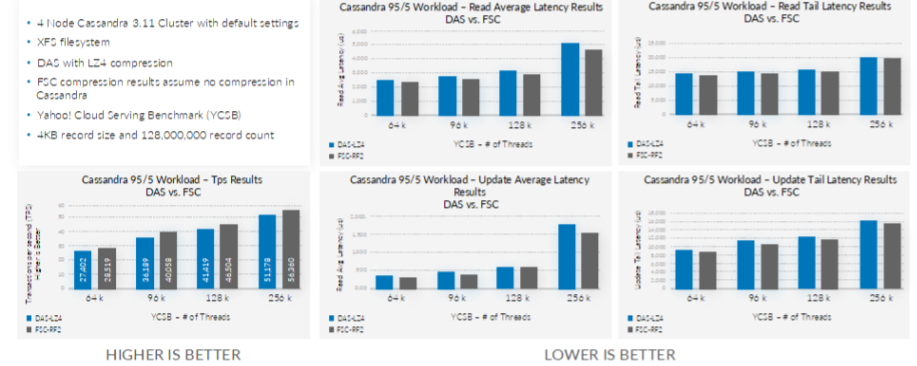}
    \caption{Cassandra DBMS: Fungible FSC vs. DAS.}
    \label{fig:dpuperf}
\end{figure}

Figure~\ref{fig:dpuperf} shows the performance of Fungible’s disaggregated storage system versus server attached storage, also known as Direct Attached Storage (DAS), for a database workload and shows that performance is indistinguishable.

Additional details on Fungible’s storage offering using DPUs can be found in~\cite{jai20221,jai20222}.

%% file: liqid.tex
Composable Disaggregated Infrastructure (CDI) solutions add a software component to hardware disaggregation.
CDI solutions consist of three parts: the disaggregated hardware components, 
a fabric that connects the disaggregated hardware components, and software (sometimes referred to as a composer) that allows for dynamically configuring the hardware components to create hardware infrastructure that precisely matches workload needs.

Instead of physically ordering hardware infrastructure with the required cores, memory, GPUs and Storage, CDI can dynamically compose and deploy such infrastructure in minutes. 

Liqid CDI disaggregates a datacenter's infrastructure using PCIe-deployed devices.
This includes GPU’s, FPGA’s, SSD’s, and Storage Class Memory.
NIC’s are not installed in the server chassis.
Instead, they are disaggregated, and placed into external PCIe enclosures, called expansion chassis.

An organization can have as many expansion chassis as are required to hold their storage, accelerators, and/or networking resources. Liqid is vendor agnostic enabling customers to choose what resources they compose. 
%Liqid’s hardware compatibility list is always growing, visit
Visit 
\url{https://www.liqid.com/resources/all} for details.

Liqid’s composable fabric switch then interconnects all the resources to be composed, providing every composable resource direct access to each other. For connections between chassis and racks to be transparent to workloads, Liqid leverages high bandwidth technology, including PCIe (Gen 4 and Gen 3). While Liqid also supports Ethernet and InfiniBand (Eth/IB) fabrics, they are not covered in the scope of this document. 

Organizations choose the fabric type that best meets their composability requirements and can even create multi-fabric environments that support them all. 
All expansion chassis support PCIe and a subset support either PCIe or Eth/IB connectivity. Compute resources (servers/blades) are connected to the fabric via PCIe HBA and/ or 100GbE network cards. All composable resources connected to high-speed fabric switches, either PCIe or Ethernet. 

Once resources are disaggregated and connected over distributed fabric(s), the data center has essentially been flattened and turned into massive computing, accelerator, and storage pools. At this point, Liqid Matrix™ composable software is used to create bare metal servers composed of resources tuned to meet any workload need, in seconds.

Liqid Matrix software lives on the fabric, allowing IT to configure and deploy servers that meet explicit workload requirements in seconds via software without worrying if a server can physically support its GPU and or storage resources. If demand increases, add more resources on-demand. As compute needs evolve, unused resources can be reclaimed for use by other applications.  Liqid Matrix CDI software enables organizations to:
\begin{itemize}
    \item Accelerate time-to-results with right-sized systems, deployed real-time via software;
    \item Adapt in real-time to evolving business needs with a zerotouch, change-ready agility;
    \item Drive new levels of efficiency with superior resource utilization and an as-a-service approach to infrastructure;
    \item Save on capital and operational expenses while providing a dynamic, disaggregated infrastructure that is part of a more sustainable, software-defined data center ecosystem.
\end{itemize}

Liqid has seen adoption for the following use cases:
\begin{itemize}

\item AI+ML \& data science: Accelerate time-to-research for scientists by enabling them to tailor systems that meet challenging workload needs in seconds, rather than forcing them to manually configure servers and accelerators. %Liqid AI-ready infrastructure allows researchers and data scientists to quickly deploy precise configurations to successfully train and deploy their machine learning models.

\item HPC: Get answers to today’s most urgent questions faster by composing previously impossible server configurations for HPC. Quickly deploy systems with precise amounts of GPU, accelerator, storage, and networking, eliminating the need to manually install or remove components from the server chassis.

\item End user computing: Meet the most demanding virtual desktop infrastructure (VDI) requirements with Liqid CDI. Compose only what’s needed to meet today’s desktop requirements and then scale GPU resources up or down via software as workload demands dictate.

\item Server virtualization: Extend the flexibility of virtualization to VMware ESX host servers with Liqid CDI. Quickly compose bare-metal host servers that meet precise workload requirements all via the Liqid vCenter Plug-in. No longer is GPU performance capacity limited by what can fit in a server. Increase VM and workload density with Liqid.

\item Edge computing: Cameras, sensors, and cell phones will continue to create vast amounts of data; edge compute is increasingly needed to process that data quickly. Liqid composability creates flexible configurations that make edge deployments a reality. Liqid is uniquely suited to address common edge challenges including limited power, floor space, cooling and human access.
\end{itemize}

%After deploying and scaling servers the same way for more than two decades, 
Software-defined CDI enables IT organizations to reap cloud-like speed and flexibility in their own core and edge infrastructure. %Liqid will continue to develop our software and compatibilities in a way that delivers maximum value, and exceeds modern workloads demands. 

%% file: memory.tex
Stateful on-line applications maintain a large amount of data and require fast processing times.
Examples include interactive games, fraud detection, and social networking, among others. 
These applications cache their data in memory to provide fast response time.
If main memory were free, these applications would cache all their data in memory.
However, memory is not free.
Moreover, each server in the cloud has a limited amount of memory to offer an application.
This limitation is exacerbated during peak system load of a stateful service when it requires additional memory to meet its Service Level Objective (SLO).

Today’s data centers are awash in unused main memory~\cite{redy2021,compucache2022}.  By some estimates, more than 50\% of data center memory is either un-allocated or unused at any given time~\cite{gu2017,legoos}.
One reason for this is the over-provisioning a virtual machine (VM) to handle the occasional VM resize operation without requiring a VM migration.
Another is external fragmentation when VMs allocated on a host machine leave insufficient resources to allocate another VM of a useful size.

A disaggregated hardware platform provides fast networking to enable data management systems, DMSs\footnote{A DMS includes traditional relational database management systems, key-value stores, document stores, etc.}, to access remote memory with acceptable latency, enhancing overall memory utilization.
In~\cite{redy2021}, we present an elastic memory management system named Redy.  
Redy tunes RDMA in a particular deployment to satisfy a user-provided SLO and minimize resource cost.  

A future research direction is how to size the server-local memory cache to satisfy the required latency and throughput of a workload.  This depends on the cache miss rate as a function of cache size and the latency of servicing a cache miss.
The former depends on the degree of skew of references to data which in turn dictates the placement of data.
The latter depends on the type of storage that services the miss.
Example storage include remote memory, server-local disk, or cloud storage.
This research direction may benefit from open industry standards such as Compute Express Link (CXL) for CPU-to-memory connections.
CXL is designed for high performance data center computers and provides cache-coherent protocols for accessing system memory (CXL.cache) and device memory (CXL.mem), and block input/output protocol (CXL.io).

%% file: alloydb.tex
%\begin{enumerate}
%\item Disaggregation of storage engine from primary and read pools.
%\item Disaggregated storage engine. 
%
%\item Pluggable columnar engine. 
%\end{enumerate}

Today’s applications have high demand on database performance and stringent service level agreement (SLA) requirements. 
An SLA may require a database to provide high throughput with a 90$^{th}$ percentile response time less than a given threshold, e.g., 100 milliseconds. Applications also have a high velocity. They want to perform complex analytics on fresh data to provide the latest business insights.

\begin{figure}[!ht]
    \centering
    \includegraphics[width=0.9\linewidth]{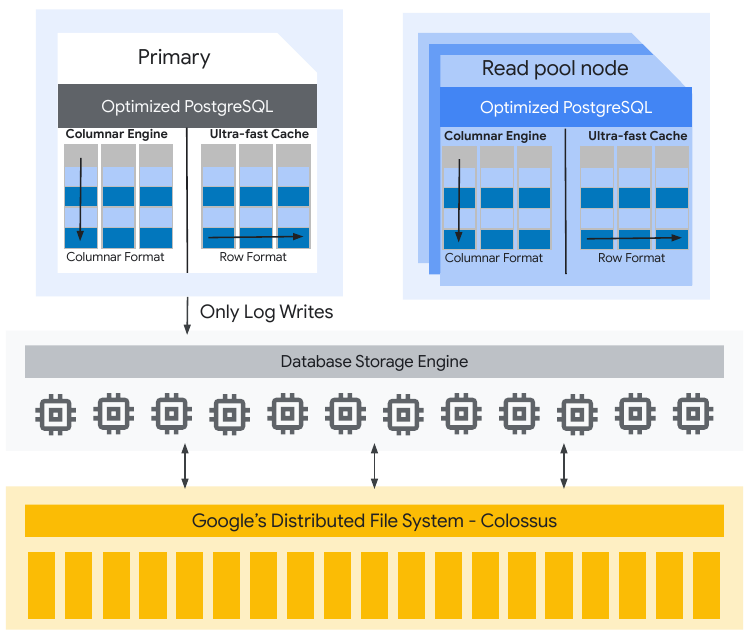}
    \caption{AlloyDB architecture.}
    \label{fig:alloydb-arch}
\end{figure}

AlloyDB is a new enterprise grade SQL database product that aims at meeting those application demands. It combines PostgreSQL with compute-storage disaggregation, read pools for horizontal scalability, and HTAP support. 
Figure~\ref{fig:alloydb-arch} shows the architecture of AlloyDB, with the primary database instance, a set of read pools, and an intelligent, distributed storage engine as the key building blocks.
AlloyDB  disaggregates the primary and read pools from the database storage engine and enables each layer to scale independently of the others. 
An AlloyDB cluster consists of one primary and multiple read pools. 
A read pool consists of multiple read replicas. 
The storage engine persists data in the distributed file system, Colossus. 

Read pools isolate performance for different workloads of an application. An application may categorize its workloads and issue queries from the same category to the same read pool. For example, a wholesale retailer may create three read pools for its web sales, store sales, and catalog sales. This ensures that read queries for web sales are isolated from the queries for store sales. 

Read pools also provide horizontal scalability. 
A read pool balances the load across its replicas. 
An application may scale the number of replicas in a read pool dynamically based on the system load. 
The scaling is elastic and does not require moving data in the storage engine.
When the system load becomes high, an application may create additional replicas to shed load. 
It may reduce the number of replicas in a pool when the load dials down. 

To provide fast query response time for both transactional and analytical queries, the primary and replicas employ an ultra-fast row cache and a pluggable columnar engine. 
The row cache caches the working set in row-oriented format while the columnar engine caches them in column-oriented format. 
Columnar engine uses single instruction, multiple data (SIMD) vectorization to facilitate query processing.
It also monitors the workload and automatically columnarizes data to maximize performance. 
The primary and replicas process a query using the columnar engine, the cached data in row format, or a hybrid of two.

\begin{figure}[!ht]
    \centering
    \includegraphics[width=0.9\linewidth]{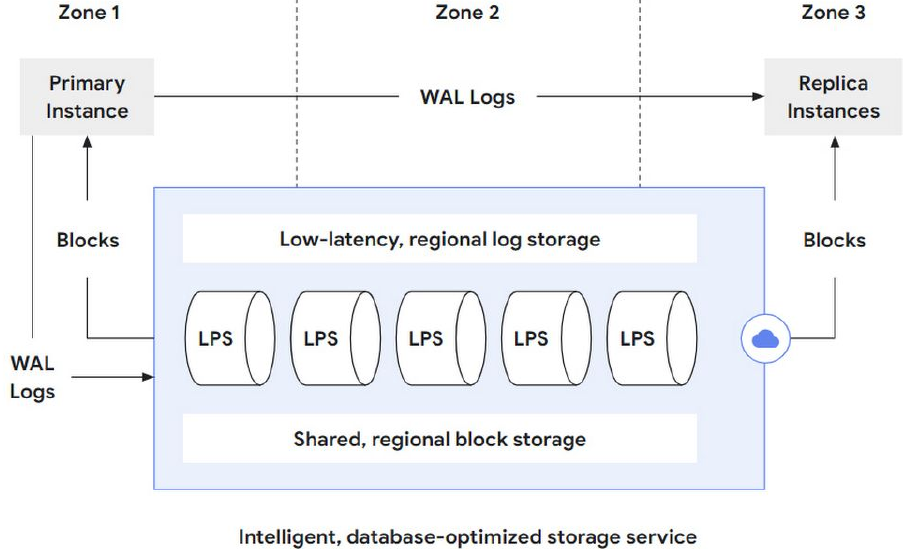}
    \caption{AlloyDB storage engine.}
    \label{fig:alloydb-storage}
\end{figure}

AlloyDB's storage engine is further disaggregated into several components to provide high performance. It also decouples durability from availability, see Figure~\ref{fig:alloydb-storage}. 
AlloyDB stores log records durably in the regional log storage and the database in the regional block storage, Colossus.
The log storage optimizes for append-only log operations to reduce the transaction commit latency. 
To ensure all blocks are readily available for primary and replicas, the storage engine uses multiple log processing servers (LPS) that ingest log records and materialize blocks from log records continuously. 
LPS materializes the blocks in the same zones as the primary and replicas. 
They are purely compute-attached to a shared regional storage and can flexibly scale without needing to copy any data. 
Internally, AlloyDB scales up the number of LPSs when the system load is high to provide consistent performance. 
It scales down during low system load to reduce cost.
The storage engine also handles the backup operations completely and does not impact the performance and resources of the compute layer.

AlloyDB also adopts a component-based architecture. 
It offers rich functionalities through multiple extensions to PostgreSQL. 
For example, the columnar engine extension plugs into the query optimizer and query execution to facilitate query processing. 
The database advisor extension provides physical database design recommendations based on the workload, e.g., indexes, and the query insights extension offers observability into the query performance. 
AlloyDB also supports widely used PostgreSQL extensions from the open-source community.

%% file: rockset.tex
%\begin{enumerate}
%    \item  Real-time databases demand disaggregation
%    \item The Aggregator Leaf Tailer (ALT) disaggregated architecture
%    \item Disaggregation of compaction cpu in RocksDB in the cloud
%\end{enumerate}

\subsubsection {Real-time databases demand disaggregation:}

%\paragraph{}
%In this section, we discuss the need of a 
This section describes a disaggregated architecture for %powering 
realtime databases. We examine, in detail, \href{http://rockset.com}{Rockset}, which is a real-time analytics database service for processing low latency, highly concurrent analytical queries at scale. 
Rockset builds a \href{https://rockset.com/blog/how-rocksets-converged-index-powers-real-time-analytics/}{Converged Index™} on structured and semi-structured data from OLTP databases, streams and lakes in real-time and exposes a RESTful SQL interface. We first discuss the requirements of a disaggregated architecture for powering realtime data applications. We then explain why the Aggregator Leaf Architecture (ALT) is able to power realtime data applications.  Finally, we present how Rockset disaggregates the RocksDB architecture to separate compute from storage.

%\paragraph{}
%If you've ever ordered food online and tracked it, you've used a data application with embedded real-time analytics. If you've ever used the Facebook or the Linkedin Newsfeed, you've seen real-time analytics in action. These are realtime applications that are powered by a realtime database. A Real-time database supports fast queries on fresh data. Fast queries means sub-second query responses. 
Many applications are powered by realtime databases.  Examples include food ordered and tracked online, Facebook or LinkedIn’s newsfeed, and others.  These applications require real-time analytics to provide interactive user requests with fresh data.  
This means fast queries with sub-second response times.
New events are streamed into the database at hundreds of megabytes a second. Fresh data means new data becomes visible to queries within a few seconds of it being updated. The update rate to the database is inherently bursty in nature and, at the same time, these realtime data applications demand that queries are not impacted by bursty writes. 
This motivates processing of the writes to be separated from the processing of reads.
This is the primary reason why disaggregation is key to powering realtime databases. 
Aggregator Leaf Tailer (ALT) is the disaggregated data architecture favored by web-scale companies, like Facebook, LinkedIn, and Google, for its efficiency and scalability for powering realtime applications.

\begin{figure}[!ht]
    \centering
    \includegraphics[width=0.9\linewidth]{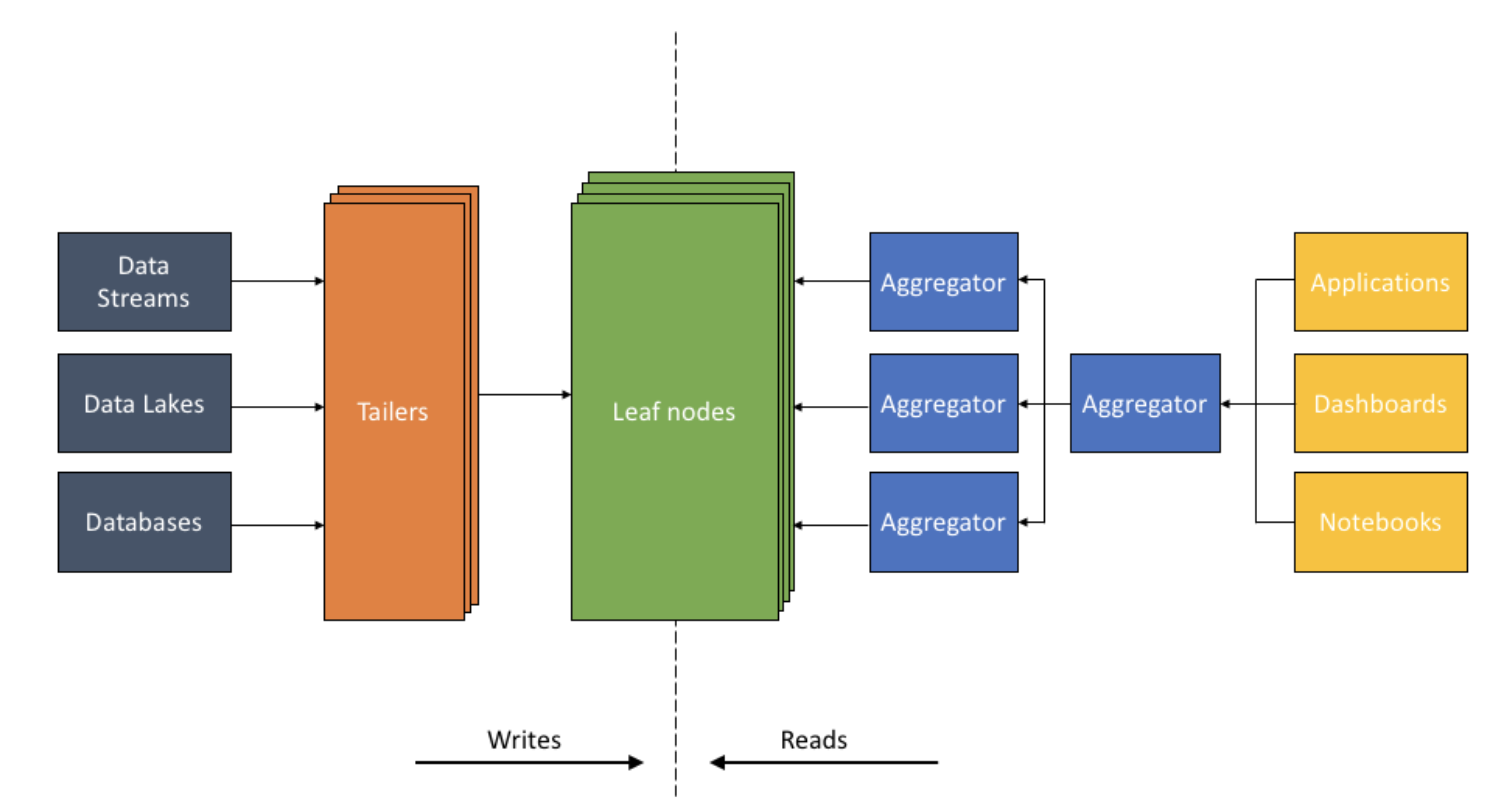}
    \caption{Aggregator Leaf Tailer Architecture (ALT)}
    \label{fig:ALT}
\end{figure}

\subsubsection {The Aggregator Leaf Tailer disaggregated architecture (ALT):}

%\paragraph{}
The \href{https://rockset.com/blog/aggregator-leaf-tailer-an-architecture-for-live-analytics-on-event-streams/}{ALT architecture} \cite{alt}  is a design pattern for realtime databases, see Figure~\ref{fig:ALT}. 
This architecture facilitates a three way disaggregation among
the compute required for writes,
the compute required for reads, and
the storage required to store and retrieve the data.  The salient features of the ALT architecture include the following.
%\begin{enumerate}
% \item The compute required for writes,
% \item The compute required for reads, and
% \item The storage required to store and retrieve the data.
%\end{enumerate}
%\noindent 
%\begin {enumerate}
% \item 
First, the Tailer pulls new incoming data from a static or streaming source into an indexing engine. Its job is to fetch from all data sources, be it a data lake, like S3, or a dynamic source, like Kafka or Kinesis.
Second, the Leaf is a powerful indexing engine. It indexes all data as it arrives via the Tailer. The indexing component builds multiple types of indexes—inverted, columnar, document, geo, and many others—on the fields of a data set. %The goal of indexing is to make a query that references any data field fast.
Its indexes expedite processing of queries that reference a data field. 
Third, the scalable Aggregator tier is designed to deliver low-latency aggregations, be it columnar aggregations, joins, relevance sorting, or grouping. The Aggregators leverage indexing so efficiently that complex logic typically executed by data-pipeline software in other architectures can be executed on the fly as a part of the query.
%\end {enumerate}

%\paragraph{} 
The ALT architecture enables the application  developer to run low-latency queries on raw data sets with minor prior transformation. A large portion of the data transformation process occurs as a part of the query itself.  There are three reasons why this is possible.
%\begin{enumerate}
%\item 
First, indexing is critical to making queries fast. The Leaves, see Figure~\ref{fig:ALT}, maintain a variety of indexes concurrently, so that relevant data can be accessed quickly regardless of the type of query—aggregation, key-value, time series, or search. Every document and field is indexed, including both value and type of each field, resulting in fast query performance that allows significantly more complex data processing to be inserted into queries.

Second, Queries are distributed across a scalable Aggregator tier. The ability to scale the number of Aggregators, which provide compute and memory resources, allows compute power to be concentrated on any complex processing executed on the fly.

Third, the Tailer, Leaf, and Aggregator run as discrete microservices in a disaggregated manner. Each Tailer, Leaf, or Aggregator tier can be independently scaled up and down as needed. The system scales Tailers when there is more data to ingest, scales Leaves when data size grows, and scales Aggregators when the number or complexity of queries increases. This independent scalability allows the system to bring significant resources to bear on complex queries when needed, while making it cost-effective to do so.
%\end {enumerate}

%\paragraph {}
The ALT architecture has been in existence for almost a decade, employed mostly on high-volume real-time data systems. \href{http://engineering.fb.com/2015/03/10/production-engineering/serving-facebook-multifeed-efficiency-performance-gains-through-redesign/}{Facebook's Multifeed Architecture} \cite{facebookalt} has been using the ALT methodology since 2010, backed by the open-source \href{http://rocksdb.org}{RocksDB} engine, which allows large data sets to be indexed efficiently.
LinkedIn’s \href{https://engineering.linkedin.com/blog/2016/03/followfeed--linkedin-s-feed-made-faster-and-smarter}{FollowFeed} \cite{linkedinfollow} was redesigned in 2016 to use the ALT architecture. Their previous architecture used a pre-materialization approach, also called fan-out-on-write, where results were precomputed and made available for simple lookup queries. LinkedIn's new ALT architecture uses a query on demand or fan-out-on-read model using RocksDB indexing instead of Lucene indexing. Much of the computation is done on the fly, allowing greater speed and flexibility for developers in this approach. Rockset uses RocksDB as a foundational data store and implements the ALT architecture \cite{rocksetwhitepaper} in a cloud service.

\subsubsection {Disaggregation of compaction CPU in RocksDB in the cloud:}

%\paragraph {}
Rockset uses RocksDB-Cloud as one of the building blocks of its distributed Converged Index. Rockset is designed with cloud-native principles, and one of the primary design principles of a cloud-native database is to separate compute from storage.
Below, we describe how Rockset extends RocksDB-Cloud to realize this separation.  
%We discuss how Rockset extended RocksDB-Cloud to have a clean separation of its storage needs and its compute needs. 
This is open source software and may be adopted by other realtime databases. 

\begin{figure}[!ht]
    \centering
    \includegraphics[width=0.9\linewidth]{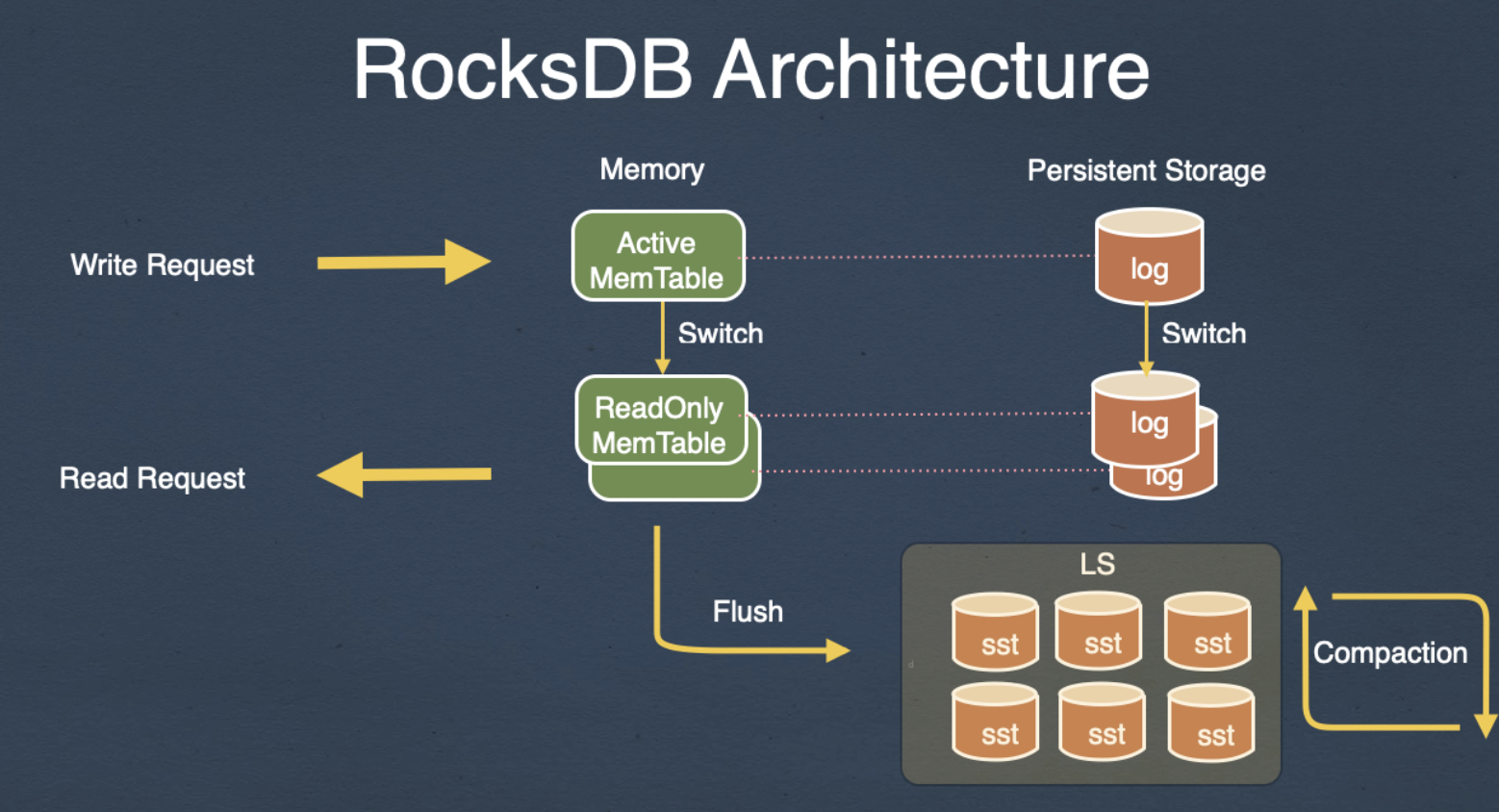}
    \caption{RocksDB}
    \label{fig:RocksDB}
\end{figure}

%\paragraph{}
RocksDB-Cloud stores data in locally attached SSD or spinning disks. %The SSD or the spinning disk provides the storage needed to store the data that it serves. 
New writes to RocksDB-Cloud are written to an in-memory memtable.
Once the memtable is full, it is flushed to a new SST file in the storage.
Being an LSM storage engine, a set of background threads are used for compaction.  Compaction is a process of combining a set of SST files and generating new SST files with overwritten keys and deleted keys purged from the new files.
Compaction is a compute intensive task.  
%Compaction requires a lot of compute resources. 
It requires more resources with a higher rate of writes to the database, enabling the system to keep up with the new writes.  See Figure~\ref{fig:RocksDB}. %the more compute resources are required for compaction, because the system is stable only when compaction keeps up with new writes to the database.

In a typical RocksDB-based system using a shared-nothing architecture~\cite{DEWI90,stone86}, compaction occurs on CPUs that are local on the server that also hosts the storage. In this case, compute and storage are not disaggregated. Hence, if the write rate increases while the total size of the database remains unchanged, the system provisions more servers to process writes.  It spreads the data across these additional servers and uses their compute resources to keep up with the compaction load.
This shared-nothing approach suffers from the following two limitations: 
%\begin{enumerate}
% \item 
First, Re-organizing data across additional servers is not instantaneous.  If the workload changes during the re-organization process then the system may not benefit from the additional servers.
% reading your data into more servers is not instantaneous because you have to copy a lot of data to do so. This means you cannot react quickly to a fast-changing workload.
% \item 
Second, The utilization of storage capacity is lowered because the database size did not change.  However, it is spread across additional servers with more storage.  This lowers price-to-performance ratio due to unused storage on the servers.
 %The storage capacity utilization on each of your servers becomes very low because you are spreading out your data to more servers. You lose out on the price-to-performance ratio because of all the unused storage on your servers.
%\end {enumerate}

Next, we describe how Rockset addresses these two limitations by separating compute from storage using
%\paragraph{}
disaggregated RocksDB-Cloud.
The primary reason why RocksDB-Cloud is suitable for separating out compaction compute and storage is because it is an LSM storage engine. Unlike a B-Tree database, RocksDB-Cloud never updates an SST file once it is created. This means that all the SST files in the entire system are read-only except the miniscule portion of data in the active memtable. RocksDB-Cloud persists all SST files in a cloud storage object store such as Amazon S3.
These cloud objects are safely accessible from all the servers because they are read-only.
%Thus, if a RocksDB-Cloud server A encapsulates a compaction job with its set of cloud objects and then sends the request to a remote stateless server B—and that server B can fetch the relevant objects from the cloud store, do the compaction, produce a set of output SST files which are written back to the cloud object store, and then communicate that information back to server A—we have essentially separated out the storage (which resides in server A) from the compaction compute (which resides in server B). Server A has the storage and while server B has no permanent storage but only the compute needed for compaction. This is disaggregation as its best!
Thus, a RocksDB-Cloud server A may encapsulate a compaction job with its set of cloud objects and send the request to a remote stateless server B.
Server B fetches the relevant objects from the cloud store, compacts them, writes a set of output SST files back to the cloud object store, 
%produces a set of output SST files, writes these SST files back to the cloud object store, 
and notifies server A that the compaction job is complete.  
In essence, %we have separated 
the storage (which resides in server A) is separated from the compaction compute (which resides in server B). Server A has the storage and while server B has no permanent storage but only the compute needed for compaction.
This disaggregation is superior to the shared-nothing approach by eliminating its two limitations.

%% file: novalsm.tex
\begin{figure}[!ht]
    \centering
    \includegraphics[width=0.9\linewidth]{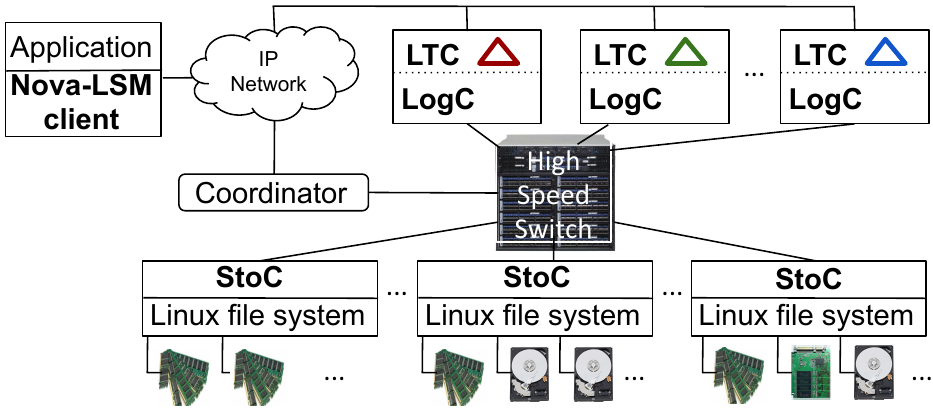}
    \caption{Architecture.}
    \label{fig:novalsm-arch}
\end{figure}

%\begin{enumerate}
%    \item Separation of storage from processing
%    \item Architecture using RDMA
%    \item Comparison graph
%\end{enumerate}

Nova-LSM is a distributed LSM-tree key-value store that disaggregates storage from processing~\cite{novalsm-sigmod}.
Figure~\ref{fig:novalsm-arch} shows its architecture, consisting of LSM-tree components (LTC), logging components (LogC), and storage components (StoC).
These components are connected using a high speed RDMA network.
Application data is range partitioned across LTCs and each LTC is assigned several ranges. 
An LTC maintains one LSM-tree for each of its assigned ranges and processes application requests using these trees. 
LogC maintains log records of a LSM-tree and is integrated into LTC. 
It generates log records when processing writes. 
It also fetches log records to recover a LSM-tree.
StoC stores, retrieves, and manages blocks. 
A StoC may consist of main memory (DRAM), non-volatile memory, disk, or a hierarchy of these storage devices. 
It leverages one-sided RDMA read/write primitives to provide high performance. 

A StoC may implement compaction as data is shared across all StoCs. 
%Data is shared across all StoCs. 
An LTC may write data to any StoC.
%To minimize storage hotspots, 
Each write request uses power-of-d to dynamically selects the fastest StoC.
The coordinator maintains the assignment of ranges to LTCs to balance load. 
In~\cite{novalsm-sigmod}, we present experimental results showing Nova-LSM provides 10x higher throughput than RocksDB~\cite{rocksdb} and LevelDB~\cite{leveldb}.% for various workloads.
%This includes faster average and 99\% response times.

\begin{figure}
%\begin{subfigure}[t]{0.5\textwidth}
\centering
\includegraphics[width=\textwidth]{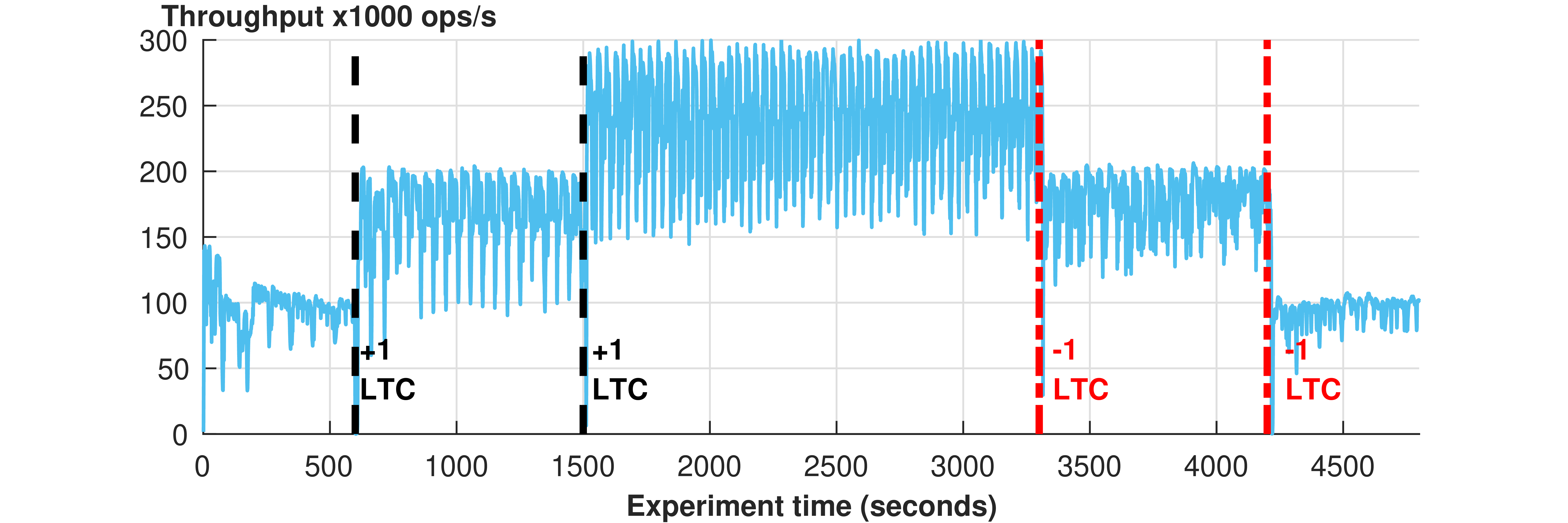}%png}%eps}
\caption{Nova-LSM is elastic.  LTCs and StoCs may scale independently.}
\label{fig:eval:nova-elastic-sw50}
%\end{subfigure}
\end{figure}

Nova-LSM is elastic.
It may scale its number of StoCs and LTCs dynamically 
%and independently 
based on system load.
When LTCs are fully utilizes, Nova-LSM may construct additional LTCs to shed load without the need to move data across StoCs.  
A new LTC is assigned one or more ranges and constructs its LSM-tree metadata to process client requests referencing its ranges. 
The LSM-tree reads data stored in a StoC.
Figure~\ref{fig:eval:nova-elastic-sw50} shows the throughput of a system as we increase and decrease the number of LTCs.  The starting configuration consists of 1 LTC and 13 StoCs.  Its peak throughput is 100,000 operations per second with the CPU cores of 1 LTC fully utilized.  An increase in the system load motivates an increase in the number of LTCs, causing the peak throughput to increase linearly as the new LTCs process requests and reduce the load on the bottleneck LTC.
As the system load decreases, the number of LTCs is reduced by removing one LTC at a time and re-assigning its assigned ranges to the other LTCs.  This causes system throughput to drop back to 100,000 operations per second with 1 LTC.

%% file: future.tex
Disaggregated database management systems are an emerging research topic that raise many interesting questions.  For example, given a workload, what is an online framework to assemble a DBMS using microservices?  
What hardware and software co-designs maximize its efficiency?  
How does one verify the correctness of a composition?
Is it possible for a system to learn patterns that maximize efficiency, enabling it to incorporate new hardware and services seamlessly? 
These and other research questions shape the future of disaggregated DBMSs.